# HALF METALLIC RESPONSE OF MANGANITE FILMS AT ROOM TEMPERATURE FROM SPIN-POLARIZED SCANNING TUNNELING MICROSCOPY

M. Cavallini, F. Biscarini, V. Dediu, P. Nozar, C. Taliani, R. Zamboni.

*Consiglio Nazionale delle Ricerche, Istituto per lo Studio dei Materiali Nanostrutturati - sezione Bologna, via P. Gobetti 101, I-40129 Bologna, Italy*

The ferromagnetic phase in $La_{0.7}Sr_{0.3}MnO_3$ manganite is completely spin polarized (100%) at room temperature. This is demonstrated on high quality epitaxial films by Spin-Polarized Scanning Tunneling Microscopy. The films consist of a highly homogeneous ferromagnetic (FM) phase in which minority paramagnetic (PM) defects are embedded (less than 1% for best films). PM defects exhibit featureless tunneling spectroscopic characteristics and an insulating-like conductance. FM phase exhibit metallic behavior and strongly nonlinear characteristics. Deconvolution of the spectroscopic curves for the FM phase reveals the totally spin-splitted $e_g t_{2g}$ manganite bands whose shape and width are in an excellent agreement with optical and photoemission spectroscopy data. These results promote the manganites as efficient spin polarized injectors at ambient conditions.

Ferromagnetic perovskite manganites $A_{1-x}B_xMnO_{3+\delta}$ (A is La, Nd, Pr *etc.*; B is Sr, Ca, Ba *etc.*) are characterized by colossal negative magnetoresistance (CMR) and by a high degree of spin polarization (SP) of the charge carriers. This feature makes manganites very attractive as spin-polarized injectors for both fundamental investigations and applications in novel devices for spintronics and quantum computing.

Both CMR and SP in manganites can be qualitatively explained by the double exchange mechanism, although for a quantitative description, charge/spin–lattice interactions and Jahn-Teller effect have to be taken into consideration[1]. The strong Hund coupling (~1 eV) favors the hopping of (1-x) $e_g$ *d*-electrons between $Mn^{3+}$ and $Mn^{4+}$ when their spin is parallel to the S=3/2 total spin of three $t_{2g}$ *d*-electrons localized at manganese sites. Below the Curie temperature ($T_c$) the delocalization of $e_g$ electrons in a narrow metallic band (metal-insulator transition) coincides with a parallel (spin-up) ordering of all $e_g$ and $t_{2g}$ electrons giving rise to a ferromagnetic state. In this simple model the spin-down band is empty and separated from the spin-up band by an energy of the order of Hund coupling.

The optimally doped $La_{1-x}Sr_xMnO_{3+\delta}$ (x=0.3, $\delta$=0) is metallic and ferromagnetic below Curie temperature $T_c \sim$ 370 K being characterized by rhombohedral structural modification[2]. At low doping levels (x<0.2) the material is an orthorhombic insulator. The orthorhombic distortion, on the other hand, can be caused by over-oxygenation ($\delta$>0) of the sample[3], that induces a cation-defective composition[4] and depresses the Curie temperature.

The half metallic properties (100% SP) of the $La_{0.7}Sr_{0.3}MnO_3$ have been demonstrated for the first time by spin resolved photoemission spectroscopy[5] at 40 K. Room temperature (~0.8$T_c$) spin polarization of this manganite is not clear so far, the literature data being spread in a wide interval[6-8]. It is authors opinion that this disagreement is also strongly induced by sample's quality.

The knowledge of the exact SP of the manganites is of fundamental importance for the spintronic applications. It has been demonstrated[9] that for the diffusive regime of conductivity at the ferromagnetic metal/ semiconductor interface the injected current cannot be spin polarized unless the ferromagnetic is nearly 100% spin polarized.

In this work we present the investigation of magnetic homogeneity and spin polarized properties of the epitaxial $La_{0.7}Sr_{0.3}MnO_3$ film surfaces at room temperature by a spin polarized Scanning Tunneling Microscopy.

Epitaxial $La_{0.7}Sr_{0.3}MnO_3$ thin films have been prepared by Channel-Spark ablation on $NdGaO_3$ substrates[3]. In order to remove the over-oxygenation effects the best films were annealed at 400÷450°C in high vacuum after deposition. Such films exhibit high $T_c$ (~350÷370 K) and resistivity lower than 10 mΩcm at 300 K. The as-grown films had a lower $T_c$ (~300÷320 K) and higher resistivity. The micro-Raman analysis of the films showed that the annealed films are essentially rhombohedral (structural phase showing a ferromagnetic metallic state[3]), with a few of orthorhombic inclusions (structural phase characterized by a strong Jahn-Teller distortion[10], i.e. an insulating paramagnetic phase), while the as-grown films consist of a strong mixture (roughly 50/50%) of rhombohedral and orthorhombic phases.

In spin polarized STM the maximum transmissivity of carriers between two spin polarized electrodes (tip and sample) separated by a tunnel junction occurs for their parallel SP and drops with increasing misalignment (spin-valve effect)[11,12]. In our experiment STM was operated at room temperature. For each bias voltage $V_{bias}$, we acquired both topography z(x,y) and $dI(x,y)/dV$ (differential tunneling conductance) maps by applying a low-frequency modulation on the $V_{bias}$ while holding the feedback. We used electrochemically etched Ni wire tips[13], which have been shown to act effectively as injector of carriers with strong spin polarization[14,15].

The typical morphology of a 100-nm thick film exhibits a smooth background (less than 5nm roughness on 8 μm lengthscale) with a low density of outgrowths, whose lateral size ranges from several hundred nanometers up to a few microns (Fig. 1a,c). The $dI(x,y)/dV$ maps (Fig. 1b) which are sensitive to local conductance show the coexistence of well defined high conductance (HC) and low conductance (LC) regions. The *dI(x,y)/dV* contrast between LC and HC regions amounts to more than one order of magnitude (Fig. 1d).

The LC defects (typical dimensions of the order of few microns) cover less than 1% of the film surface in samples post-annealed in vacuum, while the as-grown films consist roughly of 50/50% mixture of HC and LC regions (similarly to the La-Ca-Mn-O samples studied earlier[16]). Both HC and LC regions are homogeneous and can be characterized by different conductance values. The variation of the conductance is due to the metallic nature of ferromagnetic HC regions and insulating nature of paramagnetic LC defects[16]. The results are in good agreement with micro-Raman studies, where a similar fraction of the surface was found to correspond to the orthorhombic paramagnetic phase.

The high homogeneity of the HC regions does not necessarily indicate the absence of magnetic domains. First, the magnetic domains could be substantially larger that the scanned area. On the other hand, for a tip-substrate separation of a few angstroms, a strong effective magnetic field should orient the SP at the tip apex parallel to the local surface magnetization preventing the domains observation. The tip-surface interaction is a well known effect in magnetic force microscopy[17]. For magnetically hard film and magnetically soft Ni tip (our case) it becomes energetically convenient to orient the closing tip domain parallel to the film local stray field.

The magnetic nature of the spectroscopy images becomes evident from the evolution of the local conductance versus $V_{bias}$. (Fig. 2) A new method for processing the spectroscopic data was used in order to obtain a higher accuracy with respect to derivatives taken on single pixels. For each $V_{bias}$, a pixel-by-pixel average of *dI(x,y)/dV* on LC and HC regions is performed. Finally, the *dI(x,y)/dV* values are normalized to $I/V_{bias}$ to yield an estimate of *d(lnI)/d(lnV)*, which could be related to the electronic density of states (DOS). It has been shown that such a procedure removes the effect of voltage and tip-sample separation distance on tunneling current[18].

The *d(lnI)/d(lnV)* curves in Fig. 2 clearly show different responses from the HC and LC regions. HC characteristics lie above the LC ones at any bias voltage. The LC curve is smooth and extrapolates to zero at $V_{bias} = 0$ indicating an insulating behavior. The HC curves on the other hand, exhibit a metallic and non-linear behavior with few distinct features: a small narrow peak at 1 V, a sharp growth at 1.3-1.5 V, and two local maxima at 1.9 and 2.6 V. A good agreement between data taken with different Ni tips on HC regions of different films was found. No difference was observed between annealed and as-deposited films, apart for the fractional coverage by LC defects.

The manganite bands deconvolution was performed by using the orientation averaged spin dependent DOS for the Ni [19] (Fig. 3). This approximation provides reasonable starting approach for the case of polycrystalline tips with an unknown crystal orientation at the apex. Even for single crystalline tips the tunneling process will never follow exactly a given crystalline axis direction, due to finite tip curvature and non zero film roughness.

The tip-sample current is calculated by taking into account two separate spin-up and spin-down channels[20]. Thus, the general formula for tunneling current[21] is modified into:

$$I_{\uparrow(\downarrow)} \cong \frac{2\pi e}{\hbar} \left(\frac{\hbar^2}{2m}\right)^2 \int_0^{eV} T_{\uparrow(\downarrow)}(E)\, N_{\uparrow(\downarrow)}(E)\, N'_{\uparrow(\downarrow)}(E-eV)\, dE \quad (1)$$

Here $T_{\uparrow(\downarrow)}(E)$ is the transmission coefficient of the tunnel contact, $N_{\uparrow(\downarrow)}(E)$, $N'_{\uparrow(\downarrow)}(E)$ are spin polarized densities of states of Ni and manganite respectively. We evaluated eqn.(1) in the approximation of $T$ independent of energy. This approximation should be also valid for the case where T is a smoothly varying function of energy. It influences mainly the intensity of the DOS features and cannot shift significantly the characteristic energies. The total current $I$ has been calculated as the sum of spin-up $I_\uparrow$ and spin-down $I_\downarrow$ currents:

$$I = I_\uparrow + I_\downarrow \quad (2)$$

The normalized differential conductance $d(lnI)/d(lnV)$ was calculated numerically, taking into account the constant current mode experimental procedure.

In a first approximation our model was restricted to the 3$d$ Mn bands, namely two-lorentzian ($e_g$ and $t_{2g}$) bands for any spin orientation, while Ni bands were represented by literature data[19] (Fig. 3). The STM method does not distinguish between $e_g$ and $t_{2g}$ symmetries, so we will keep $e_g t_{2g}$ notation for the calculated band structure. Fitting with elliptical and gaussian manganite bands showed no qualitative difference. The choice of lorentzian bands is justified however by the best data-fit agreement and has been successfully used previously for manganite DOS calculations[22].

At positive biases, electrons move from the tip to the manganite, so that only the empty states in the manganite bands are relevant. We take the Fermi energy position, the band widths, and band (spin up – spin down) splitting as fitting parameters.

The resulting band structure is represented in Fig. 4 by a solid curve. It generates the solid fitting curve for $d(lnI)/d(lnV)$ in Fig. 2 in excellent agreement with experimental data. The maximum of the normalized differential conductance curve at 1 V (Fig. 2) is caused by the overlap of the spin-up band of $La_{0.7}Sr_{0.3}MnO_3$ (tail in Fig. 4) with the 1 eV maximum in the spin-up DOS of Ni (Fig. 3). The steep increase at 1.3-1.5 V (Fig. 2) results from the overlap of the Ni spin-down peak at Fermi level and $La_{0.7}Sr_{0.3}MnO_3$ spin down band tail (Fig. 4). Finally, the two local maxima (1.9 and 2.6 V) are a consequence of the splitting of the manganite spin-down band in two subbands and the overlap of the 3 eV maximum in the Ni spin-up band with the spin-up manganite band tail.

The only sub-band with non-zero DOS at $E_F$ is the tail of the spin-up polarized band (Fig. 4), which extends roughly up to 1.5 eV (the energies are calculated from $E_F$). This band is associated with the tail of the spin up $e_g t_{2g}$ band of Mn[5, 23]. The spin down band is an overlap of two lorentzian sub-bands, and has non-zero DOS from 0.4 eV up to more than 4 eV. The band width is roughly 2 eV, and it is characterized by a maximum at 2.5 eV and a shoulder at 1.8 eV.

The first approximation provides a good fitting for the bias voltages in interval 0-2.5 V, while it deviates at higher voltages. By adding a spin unpolarized band that starts at 2.5 eV and reaches the constant DOS at 3.4 eV, the fitting procedure is further improved (blue dashed lines in Figs.2 and 4). This band can be associated with the La(5$d$) spin unpolarized band[23].

The most striking feature of the manganite band structure is 100% polarization of charge carriers at $E_F$ due to the fully developed gap in spin-down polarized DOS. We would like to draw the attention to the fact, that due to the high maximum in spin-down polarized DOS of Ni at $E_F$, our measurements are much more sensitive to the presence of non-zero spin-down DOS than spin-up DOS at $E_F$. To the best of our knowledge this is the first demonstration of the half metallic properties of manganites at room temperature. The shape and characteristic energies of derived bands are in good agreement with theoretical ab-initio calculations and optical data[23, 24]. A remarkable agreement between proposed spin down band and $t_{2g}e_g$ spin up band calculated from spin polarized photoemission should be also noticed[5].

We should notice that the adopted here approximations (energy independent tunneling coefficient, two lorentzian band, and orientation averaged DOS for Ni tip) provided the possibility to reveal only the main features of the manganite band structure. A further improvement (band fine substructure) can be obtained on one hand by using single crystalline tips of different spin polarized materials, and, on the other hand, by more sophisticated numerical approach.

An important result of the SP-STM study is the high magnetic homogeneity for the annealed films. Within the accuracy of our experimental setup, that amounts to 0.1 nA/V for $dI(x,y)/dV$ maps, no phase separation in the ferromagnetic regions was detected. This is in accordance with the dynamical mean field calculations for the FM double exchange systems of this doping level (x=0.3)[24]. On the other hand, the charge ordered manganites exhibit a distinct electronic phase separation on nanoscopic scale.

The spin polarized STM has a spatial resolution that allows us to investigate separately the ferromagnetic phase and the "non-magnetic" inclusions. This is the reason for the apparent disagreement between our results and some TM and PES data[6, 7] that indicate the $La_{0.7}Sr_{0.3}MnO_3$ surface as only partially spin polarized at room temperature. In the later case all the magnetic defects are integrated in the final result, while the defect density, on the other hand, depends strongly on preparation procedure (see above).

In conclusion, a 100% SP of La$_{0.7}$Sr$_{0.3}$MnO$_3$ film surfaces at room temperature was found. A high contrast and nanometer spatial resolution coupled with few tens-eV resolution spectroscopy information of magnetic phases in a manganite thin film was achieved by spin polarized STM. Apart from the low conductance defects (less than 1% for the best samples) no additional phase separation is detected on the dominating high conductance regions, indicating high magnetic homogeneity in La$_{0.7}$Sr$_{0.3}$MnO$_3$ down to 50 nm lengthscale. We believe these properties are common for high quality oxygen homogeneous films of different ferromagnetic manganites.

We are grateful to Erio Tosatti for critical discussions. We acknowledge also partial support to this work from EU-RTD project SPINOSA.

@ V.Dediu@ism.bo.cnr.it

Figure captions:

**Fig. 1**. STM topography (a) of a defect-rich zone of the manganite surface and corresponding *d(lnI)/d(lnV)* map (b) acquired simultaneously. Topographic line profile (c) corresponds to the red line on the topography map, while differential tunneling conductance line profile (d) corresponds to the blue line on the *d(lnI)/d(lnV)* map.

**Fig. 2**. Logarithmic derivatives *d(lnI)/d(lnV) = U/I*dI(x,y)/dV* acquired at $I = 1$ nA and different bias voltages: data are taken separately for high conductance (HC) and low conductance (LC) regions. The curves represent the results of deconvolution calculations.

**Fig. 3**. Spin polarized Ni DOS[21].

**Fig. 4**. $La_{0.7}Sr_{0.3}MnO_3$ spin polarized DOS at 300K from the deconvolution of tunneling spectroscopy data.

**Fig. 1**

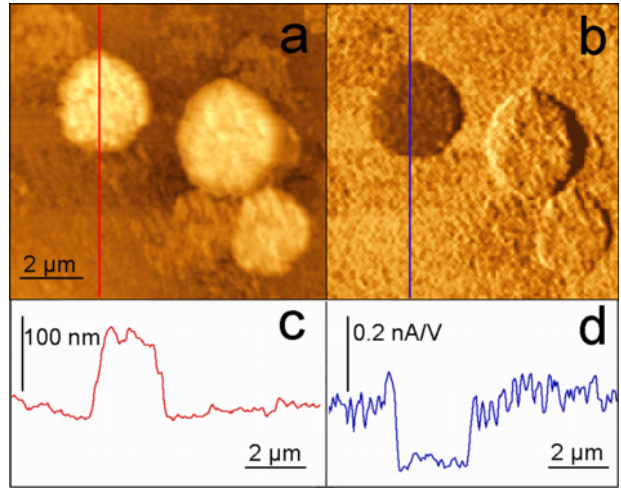

**Fig. 2**

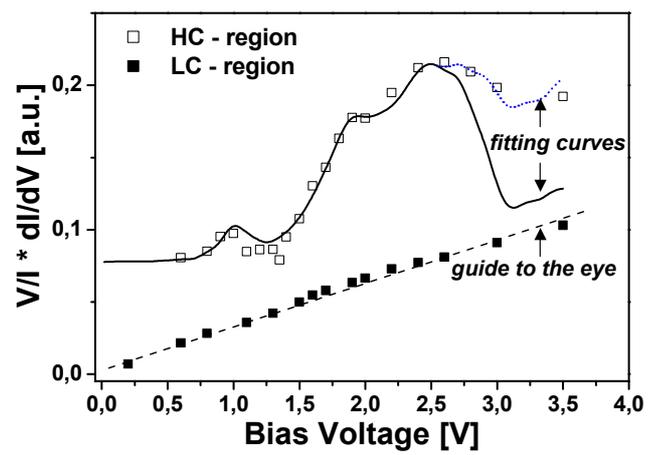

**Fig. 3**

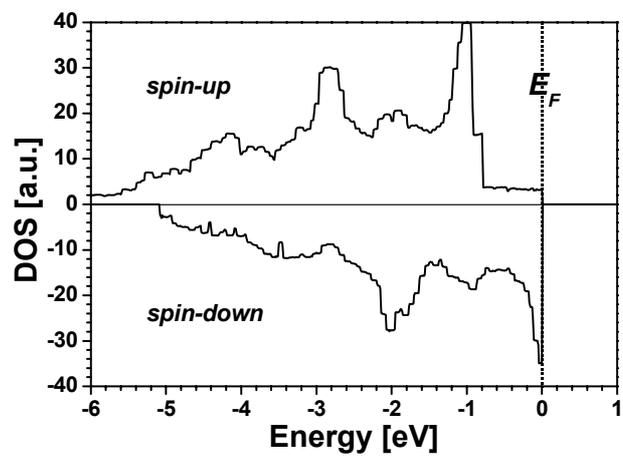

**Fig. 4**

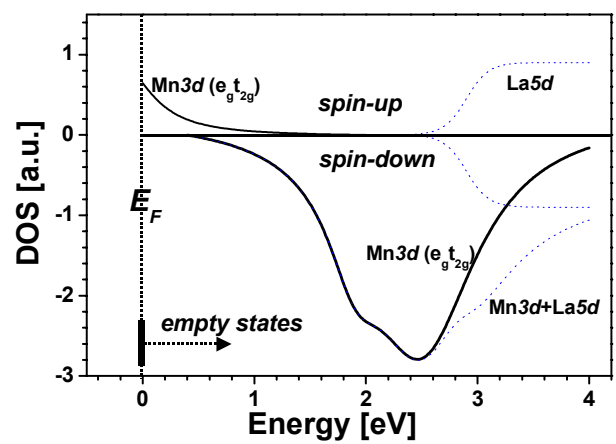